\documentclass[%
 aip,
 amsmath,amssymb,
 reprint,%
]{revtex4-1}

\usepackage{graphicx}
\usepackage{dcolumn}
\usepackage{bm}

\usepackage[utf8]{inputenc}
\usepackage[T1]{fontenc}
\usepackage{mathptmx}
\usepackage{etoolbox}

\makeatletter
\def\@email#1#2{%
 \endgroup
 \patchcmd{\titleblock@produce}
  {\frontmatter@RRAPformat}
  {\frontmatter@RRAPformat{\produce@RRAP{*#1\href{mailto:#2}{#2}}}\frontmatter@RRAPformat}
  {}{}
}%
\makeatother

\usepackage{CJK}
\usepackage{enumerate}
\usepackage{graphicx}
\usepackage{dcolumn}
\usepackage{bm}
\usepackage{graphicx}
\usepackage{bbold} 
\usepackage{graphicx}
\usepackage{dcolumn}
\usepackage[left=3cm,top=2.5cm,right=3cm,bottom=2.5cm]{geometry}
\usepackage{array}
\usepackage{epsfig}
\usepackage{wrapfig}
\usepackage{color}
\usepackage{soul}
\usepackage{braket}
\usepackage{verbatim}
\usepackage{mathtools}
\usepackage{mathrsfs}
\usepackage{amsfonts}
\usepackage[english]{babel}
\usepackage{textcomp}
\usepackage{float}
\usepackage{comment}
\usepackage{url}
\usepackage{multirow}

\excludecomment{toexclude}
\usepackage{changepage} 

\begin{document}

\begin{CJK*}{UTF8}{}
\CJKfamily{gbsn}

\title{Charge Densities and Triply-Periodic Minimal Surfaces in Crystalline Materials}

\author{Mengdi Yin (尹梦迪), Dimitri D. Vvedensky}
\affiliation{ The Blackett Laboratory, Imperial College London, London SW7 2AZ, United Kingdom}

\date{\today}
\begin{abstract}

The relationship between surfaces of constant charge density and triply-periodic minimal surfaces (TPMS) has been the subject of considerable speculation over many years.  Zero-potential surfaces generated by an electrostatic field from a distribution of point charges corresponding to a crystal provide an approximate description of the TPMS for the structure of that crystal. We have recently provided a first-principles alternative to such phenomenological comparisons based on the Vienna {\it ab initio} simulation package (VASP).  We showed that the surfaces of zero charge density calculated for the crystal structure of a material converges to the TPMS of the corresponding crystalline material. The exchange-correlation potentials are chosen for the particular material based on the benchmarking of various approximations for these potentials carried out by others. Here, we report an extension of our previous work by giving additional examples of our theory that shows the zero electron density of an equilibrium structure corresponds to a TPMS for a variety of materials and crystalline structures. We study the ground states of elemental materials that differ electronically and structurally, Na, Cu, Al, Zr, and a compound, NiTi, as well as different phases of the elemental solids that are observed by varying the different thermodynamic conditions. 

\end{abstract}


\maketitle
\end{CJK*}

\section{Introduction}

Surfaces whose area is a minimum with prescribed boundary conditions are called minimal surfaces. A typical example is the bubble film. When a bounding wire is submerged into soapy water and then pulled out of the water, a bubble film forms within the wire. The shape of the film has the least free energy and, thus, is the most stable. Since the surface free energy is proportional to the surface area, the question becomes how to find the surface whose area is minimized when the boundary is given. Such a surface is called a minimal surface.\cite{hyde96,fomenko91}

To define a minimal surface in mathematical terms, we consider the normal curvature of a surface at a given point. The normal curvature, denoted as $k_n$, is a measure of how much the surface $S$ curves in a specific direction, defined by the intersection of $S$ with a normal plane at that point. The normal curvature is a signed quantity, indicating whether the surface curves towards (positive) or away (negative) from the surface normal at the point. \cite{fomenko91,carmo16} The maximum $k_1$ and minimum $k_2\le k_1$ normal curvatures are known as the principal curvatures of $S$ and their directions are orthogonal.  From the principal curvatures we define the mean curvature $M={1\over2}(k_1+k_2)$ and the Gaussian curvature $G=k_1k_2$. 

The Gaussian and mean curvatures have greater geometrical significance than the principle curvatures. In particular, the surface is locally elliptic (bulges outward/inward) if $G>0$, locally hyperbolic (saddle-shaped) if $G<0$, locally flat if $G=0$, and locally minimal if $M=0$.  Minimal surfaces are defined as surfaces whose mean curvature vanishes everywhere. \cite{hyde96,carmo16} Thus, at every point on a minimal surface, the principal curvatures have opposite sign (resp., or both vanish), which constitutes a saddle point (resp., or a flat point).

\begin{figure*}
\centering
\includegraphics[width=0.8\linewidth]{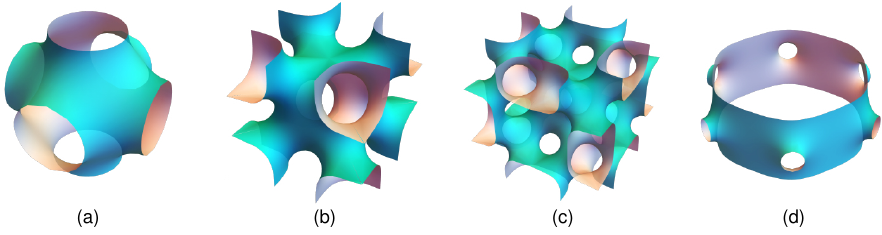}
\caption{Periodic units that repeat along three independent directions to form a TPMS: (a) P surface, (b) I-WP surface, (c) F-RD surface, (d) H$^\prime$-T surface. Table~\ref{tab1} shows the crystalline materials corresponding to these TPMS.}
\label{fig1}
 \end{figure*}

Minimal surfaces that repeat  in three independent directions are called triply-periodic minimal surfaces (TPMS).  Thus, TPMS have both zero mean curvature at every point and crystallographic symmetry.  Figure~\ref{fig1} shows examples of TPMS:~in standard notation, periodic units are shown of the Schwarz P surface, the I-WP surface, the F-RD surface, and the Schoen H$^\prime$-T surface. \cite{schoen70,schwarz90} The space groups, point groups, and genus of the TPMS studied here are compiled in Table.~\ref{tab1}. 

\begin{table*}
\caption{Symmetry groups of materials studied here with their corresponding TPMS. The abbreviations are space group (SG) and point group (PG). The genus is a topological invariant of a surface that can be interpreted as the number of holes in the surface. \label{tab1}}
\begin{ruledtabular}
\begin{tabular}{ccccccc}
Material& SG (Material) & PG (Material) & TPMS  & SG (TPMS) & PG (TPMS)& Genus\\
\hline
\vspace{-3mm}\\
Na  	& $Im\overline{3}m$ 		& $O_h$ 	    	& I-WP 		     	& $Im\overline{3}m$ 		& $O_h$ & 4 \\
Na & $P6_3/mmc$ & $D_{6h}$ & H$^\prime$-T  &  $P6/mmm$ & $D_{6h}$ & 4\\
Al   	& $Fm\overline{3}m$ 	& $O_h$	    	& F-RD 	     		& $Fm\overline{3}m$ 	& $O_h$ &6\\
Al & $Im\overline{3}m$ & $O_h$ &I-WP & $Im\overline{3}m$ & $O_h$ & 4 \\
Cu  	& $Fm\overline{3}m$ 	& $O_h$	    	& F-RD 	     		& $Fm\overline{3}m$ 	& $O_h$&6\\
Cu & $Im\overline{3}m$ & $O_h$ &I-WP & $Im\overline{3}m$ & $O_h$ & 4 \\
NiTi	& $Pm\overline{3}m$ 	& $O_h$ 	    	& P 		     		& $Pm\overline{3}m$ 	& $O_h$ &3\\
Zr  	& $Im\overline{3}m$ 		& $O_h$	    	& I-WP 		     	& $Im\overline{3}m$ 		& $O_h$ &4 \\
Zr  	& $P6_3/mmc$ 		& $D_{6h}$	& H$^\prime$-T 	& $P6/mmm$ 			& $D_{6h}$ &4 \\
Zr      &$P6/mmm$ 		& $D_{6h}$	& H$^\prime$-T 	& $P6/mmm$ 			& $D_{6h}$ &4\\
\end{tabular}
\end{ruledtabular}
\end{table*}

\begin{figure*}
\centering
\includegraphics[width=.7\linewidth]{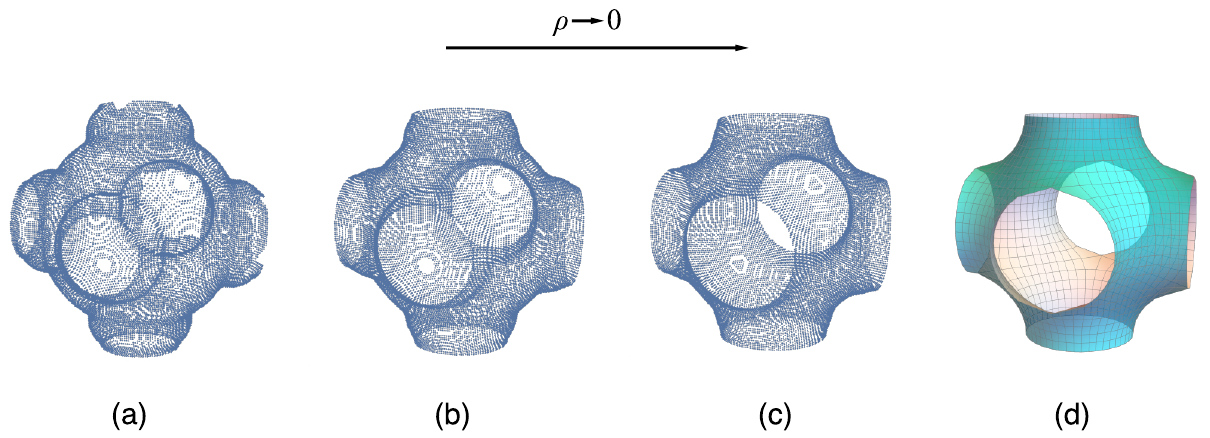}
\caption{Evolution of a surface of a given charge density in B2 NiTi. From (a) to (c), the corresponding charge density goes towards zero and the surface is going towards a P surface.}
\label{fig2}
\end{figure*}

TPMS are present in a variety of natural structures, such as the sea urchin plate \cite{jessop24}, and have been adopted when designing various materials \cite{andersson88,torquato04,han18,alketan19,lai07}, including the scaffold design of human tissue \cite{kapfer11}. However, they have not been as extensively investigated from a crystallographic perspective. Several authors \cite{schnering87,schnering91,mackay93,gandy01,gandy02} have suggested that zero equipotential surfaces (ZEP) of a suitable chosen electrostatic field from a point charge distribution provide an approximate realization of TPMS. The simple analytic forms obtained from ZEP \cite{schnering91} are an attractive presentation of TPMS and provide a {\it post hoc} justification of this approach. In particular, approximate expressions of TPMS in terms of trigonometric functions find use in 3D printing of materials \cite{restrepo17,gado24} and for the identification of atomic positions to compare with real structures. \cite{gandy01}

In previous work, \cite{yin25}, we took such largely phenomenological comparisons a step further by first using the Boltzmann equation to show that surfaces of zero charge density must correspond to minimal surfaces.  We then used calculations based on density-functional theory (DFT) to show that surfaces of zero charge density for a particular crystal converge to the TPMS for that crystalline structure, as shown in Fig.~\ref{fig2}. Here, we will provide additional examples to support this result. We show that the surfaces with vanishing charge density do indeed correspond to the TPMS of the structure corresponding to the material, including different phases of elemental solids. As expected, these charge density surfaces fit TPMS better than those based on an electrostatic distribution of charges, as shown by root-mean-square fits.

The outline of our paper is as follows. In Sec.~\ref{sec1}, we review the theories use to describe electron densities and potentials of minimal surfaces. In Sec.~\ref{sec2}, the salient features of DFT are discussed along with the approximations for exchange and correlation we use for particular materials. The parameters for our DFT calculations are summarized in Sec.~\ref{sec3}. We explain how we compare the results of our DFT calculations to the TPMS of the corresponding materials. Our results are presented in Sec.~\ref{sec4} for different types of material:~Na, Al, Cu Zr, and NiTi. In each case, surfaces near the zero of the charge density provide the best fit to the TPMS for the structure of the material. All results,  comparison between ZEP and their corresponding TPMS, and discussion over their implications for further studies are summarized in Sec.~\ref{sec6}. The Supplementary Information (SI) provides an introduction to TPMS, the data used to plot figures in the main text, and approximate expressions for several TPMS in terms of trigonometric functions obtained from ZEP.

\section{Minimal Surfaces and Electron Charge Density}
\label{sec1}

Our theory of how minimal surfaces emerge from charge density calculations is based on the Boltzmann equation of semiclassical transport theory, \cite{landau81}
\begin{equation}
{\partial f\over\partial t}+{q\bm{F}\over\hbar}\cdot\bm{\nabla}_kf+\bm{v}\cdot\bm{\nabla}_{\bm{r}}f=Q(f),
\end{equation}
where $f(\bm{r},\bm{k},t)\,d\bm{r}\,d\bm{k}$ is the probability of finding a particle in the volume $d\bm{r}\,d\bm{k}$ centered at $(\bm{r},\bm{k})$ at time $t$, $q$ is the charge on the particle, $\bm{F}$ is an external field, and $Q$ is the collision integral. The subscripts on the partial derivatives indicate the variable being differentiated. In the low-density approximation for $Q$, \cite{jungel09} the Boltzmann equation reduces to the continuity equation for the charge density $\rho$,
\begin{equation}
{\partial\rho\over\partial t}+\bm{\nabla}\cdot\bm{j}=0.
\label{eq2}
\end{equation}
where $\rho$ and current density $\bm{j}$ are
\begin{align}
\rho(\bm{r},t)&=\int f(\bm{r},\bm{k},t)\,d\bm{k},\\
\bm{j}(\bm{r},t)&=q\int\bm{v}(k)f(\bm{k},\bm{v},t)\,d\bm{k}.
\end{align}
The current density is the sum of the drift $\bm{j}_{\bm{F}}$ and diffusion $\bm{j}_D$ current densities:
\begin{equation}
\bm{j}_{\bm{F}}=q\mu\rho\bm{F},\quad \bm{j}_D=qD\bm{\nabla}\rho,
\end{equation}
where $\mu$ is the electron mobility and the Einstein relation \cite{reif65} is used for the diffusion constant $D$, which is assumed to be spatially homogeneous.

In the stationary limit, the continuity equation (\ref{eq2}) reduces to
\begin{equation}
\bm{\nabla}\cdot\bm{j}=\bm{\nabla}\cdot(\bm{j}_{\bm{F}}+\bm{j}_D)=0,
\label{eq6}
\end{equation}
with the divergences
\begin{align}
\label{eq7}
\bm{\nabla}\cdot\bm{j}_{\bm{F}}&=q\mu\bm{F}\cdot\bm{\nabla}\rho+q\mu\rho\bm{\nabla}\cdot\bm{F},\\
\bm{\nabla}\cdot\bm{j}_D&=qD\bm{\nabla}^2\rho.
\label{eq8}
\end{align}

Equations (\ref{eq6})--(\ref{eq8}) are the equations we use to derive our result. By assuming only that there is no charge accumulation near any point on a surface with $\rho=0$, we arrive that the condition that the sum of the principal curvatures vanishes:
\begin{equation}
k_1+k_2=0\, .
\label{eq9}
\end{equation}
This equation must hold at every point on the surface of $\rho=0$, which must, therefore, be a minimal surface.  The theory leading to (\ref{eq9})  is an exact result of the Boltzmann equation in the low-density limit which can be tested against DFT calculations.  This is done in Sec.~\ref{sec4}.

An alternative approach to deriving TPMS associated with crystalline materials begins with the electrostatic potential
\begin{equation}
V(\bm{r})=\int{\rho(\bm{r}^\prime)\over|\bm{r}-\bm{r}^\prime|}\,d\bm{r}^\prime,
\label{eq10}
\end{equation}
where $\rho$ is the electrostatic charge density and $\bm{r}=(x,y,z)$.  The structure factor $F_{hkl}$ is the Fourier transform of the charge density, \cite{coppens97}
\begin{equation}
F_{hkl}=V_\Omega\int_\Omega \rho(\bm{r}) e^{i\bm{K}_{hkl}\cdot\bm{r}}d\bm{r},
\end{equation}
where $V_\Omega$ is the volume of the unit cell, $\bm{K}_{hkl}$ is a reciprocal lattice vector with miller indices $(hkl)$, so
\begin{equation}
\bm{K}_{hkl}\cdot\bm{r}=2\pi(hx+ky+lz)\, .
\end{equation}
A triply-periodic charge density can be expanded as
\begin{equation}
\rho(\bm{r})={1\over V_\Omega}\sum_{{h,k,l}=-\infty}^\infty F_{hkl}e^{-i\bm{K}_{hkl}\cdot\bm{r}}.
\end{equation}
Thus, from the convolution theorem applied to (\ref{eq10}), the Fourier transform of a triply-periodic potential is
\begin{equation}
V(\bm{r})={1\over\pi V_\Omega}\sum_{hkl=-\infty}^\infty{F_{hkl} e^{-\bm{K}_{hkl}\cdot\bm{r}}\over h^2+k^2+l^2}.
\label{eq14}
\end{equation}

As an example, consider the space group $Pm\bar{3}m$ with a single type of atom.  The leading terms in the P surface approximation contain the following planes:\cite{schnering91,mackay94} 
\begin{equation}
(100), (010), (001), (\bar{1}00), (0\bar{1}0), (00\bar{1}),
\end{equation}
in which case, $h^2+k^2+l^2=1$.  The potential in (\ref{eq14}) can therefore be approximated by
\begin{equation}
V({\bm r})\approx {F_{\{100\}}\over\pi V_\Omega}\big[\cos(2\pi x)+\cos(2\pi y)+\cos(2\pi z)\big].
\end{equation}
Setting $V(\bm{r})=0$ yields the ZEP,
\begin{equation}
\cos(2\pi x)+\cos(2\pi y)+\cos(2\pi z)=0,
\label{eq17}
\end{equation}
which resembles a minimal surface, but is not a minimal surface itself.\cite{mackay94} 

\begin{figure}[t!]
\centering
\includegraphics[width=1\linewidth]{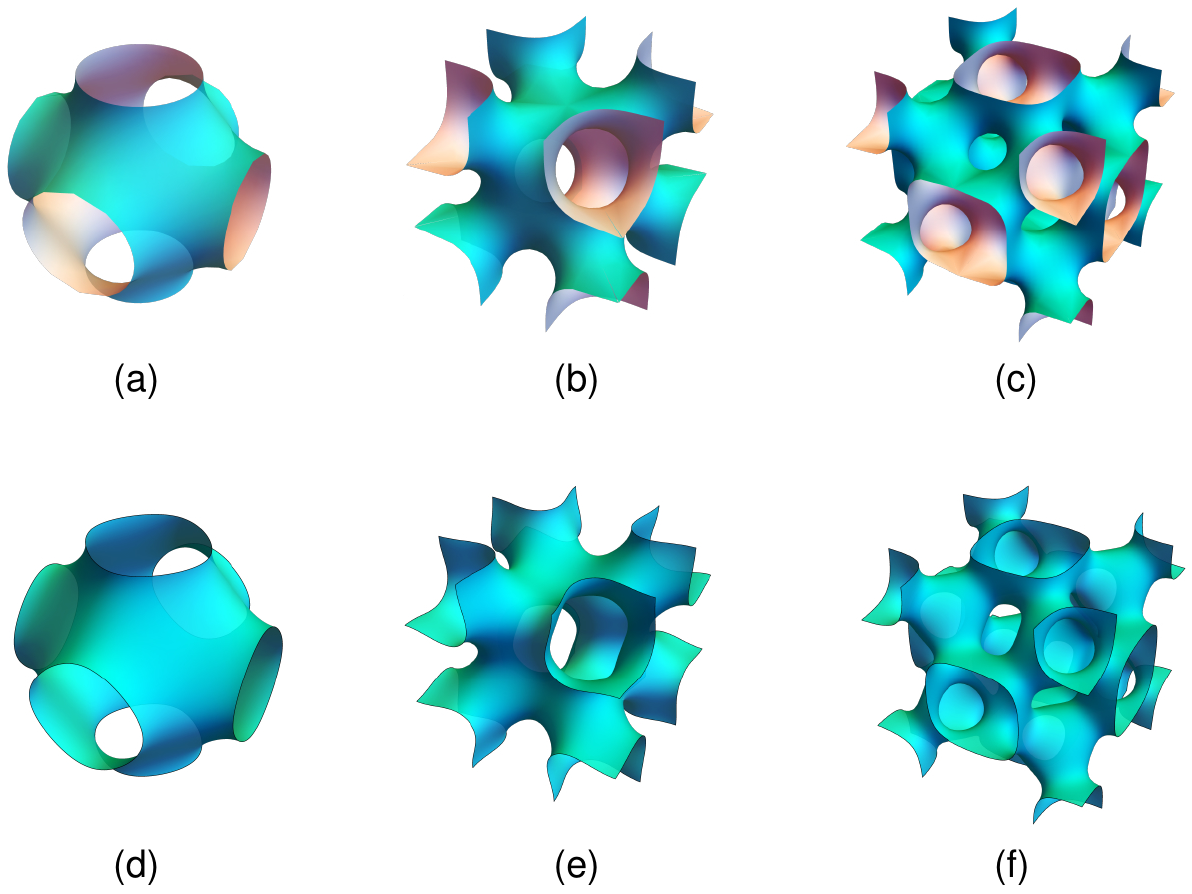}
\caption{TPMS and their corresponding ZEP. The (a) P, (b) I-WP, and (c) F-RD surfaces. ZEP, obtained from trigonometric expressions, corresponding to the TPMS in the first row: (d) P$^*$, (e) IP$_2$-J$^*$, and (f) F$_{xx}$-P$_2$F$_z$ surfaces.}
\label{fig14}
\end{figure}

Calculations based on DFT provide a much better convergence to TPMS than distributions of static charges, though the simple analytic forms \cite{schnering91} may prove useful for particular applications. Figure~\ref{fig14} compares three of the TPMS and their corresponding ZEP. Table~\ref{table3} tabulates the corresponding RMS of our DFT results and these ZEP., as described in Sec.~\ref{sec3B}. Expressions of these ZEP are given in the Supplementary Information. We have used the leading term in Fourier series to describe the ZEP, as in (\ref{eq17}), since this provides an adequate description of the minimal surface \cite{klinowski96}, though a method for obtaining corrections has been described by Mackay. \cite{mackay94} Nevertheless, the convergence of (\ref{eq14}) to the corresponding TPMS remains problematic.\cite{barnes90}

\begin{table}[b!]
	\centering
	\caption{Comparison TPMS to converged of surfaces calculated using DFT and their corresponding ZEP using RMS. The RMS is expressed in angstroms.\label{table3}}
	\begin{ruledtabular}
	\begin{tabular}{lclc}
	TPMS  &RMS (DFT) &ZEP  &RMS (ZEP) \\
		\hline
		P   &0.0171  &P$^*$  & 0.0402 \\
I-WP  &0.0568 & IP$_2$-J$^*$   & 0.1252\\
F-RD &0.0820 &F$_{xx}$-P$_2$F$_z$  &0.1537\\
	\end{tabular}	
	\end{ruledtabular}
\end{table}

\section{Density Functional Theory}
\label{sec2}

Hohenberg, Kohn and Sham \cite{hohenberg64,kohn65} formulated a theory based on the electron density that gives the solution of the Schr\"odinger equation for many-electron systems a formal basis. The idea of the Kohn--Sham--Hohenberg theorems, now called density-functional theory (DFT), is that the ``real'' electrons are replaced by ``effective'' electrons with the same total density that move as independent particles in the potential of the other electrons and the ion cores. The complex interactions between the electrons of the original system are subsumed into a universal exchange-correlation (XC) energy functional. However, DFT proves only the existence and uniqueness of this functional:~no prescription is given for its determination.  Therefore, the major challenge for DFT is choosing an appropriate approximation to the XC functional. This has led to many approximations with varying performance for materials structure and properties. 

The simplest approximation is the local density approximation (LDA).\cite{kohn65} The spatial variations of the electron charge density are assumed small, with the local charge density equal to that of a uniform electron gas. \cite{koch01} Where the LDA works well, for example, for materials consisting of $sp$ bonds, the accuracy of bond lengths and angles is a few percent. With increasing spatial variation of the charge density, we must look to corrections of the LDA.

An improvement in the accuracy of the LDA that accounts for the spatial variations of the charge density are generalized gradient approximation (GGA) functionals.\cite{ziesche98,koch01} There have been many proposed GGA functionals,\cite{perdew01}, with recent comparisons against quantum Monte Carlo calculations \cite{aouina24} for Si, NaCl, and Cu showing how the various GGA-type XC potentials performed.  

Among the most widely used GGA functionals is that of Perdew, Burke, and Ernzerhof,\cite{perdew96} known as the GGA-PBE. This is a non-empirical functional with reasonable accuracy over a wide range of systems.  Non-empirical functionals use only general rules of quantum mechanics and special limiting conditions to determine the parameters in a general form, without recourse to fitting any particular material. Such XC functional generally performed better for the tests carried out in Ref.~\onlinecite{aouina24}. 

\section{Computational Method}
\label{sec3}

\subsection{Density-Functional Calculations}

We summarized in Sec.~\ref{sec1} our theory\cite{yin25} that the surface corresponding to zero charge density is a minimal surface.  The Boltzmann equation was used to show that vanishing diffusion and drift currents along and perpendicular to the surface of zero charge density requires this surface to be minimal.  In this section, we describe how DFT calculations are used to determine the electronic structure of materials and how comparisons between surfaces of zero charge density are made.

\begin{table}[b!]
\caption{\label{table2} The phases, exchange-correlation (XC) functionals, and the number of mesh grid points used for the calculations are reported here.}
\begin{ruledtabular}
\begin{tabular}{lccc}
Material & Phase & XC & FFT Mesh  \\
\hline
\hskip5pt Na & BCC & LDA &  $140\times 140 \times 140$ \\
\hskip5pt Na & HCP & LDA & $168\times168\times168$ \\
\hskip5pt Cu & FCC & GGA-PBE & $100\times 100\times 100$ \\
\hskip5pt Cu & BCC & GGA-PBE & $120\times 120 \times 120$ \\
\hskip5pt Al & FCC & GGA-PBE & $100\times 100 \times 100$ \\
\hskip5pt Al & BCC & GGA-PBE & $120\times 120 \times 120$ \\
\hskip5pt NiTi & B2 & LDA & $120\times 120 \times 120$ \\
\hskip5pt Zr & BCC & LDA  & $112\times 112\times 112$\\
\hskip5pt Zr & Hexagonal & LDA &  $112\times 112\times 112$\\
\hskip5pt Zr & HCP & GGA-PBE &  $140\times 140\times 140$\\
\end{tabular}
\end{ruledtabular}
\end{table}

All our DFT calculations were carried out with the Vienna {\it ab initio} simulation package (VASP). \cite{kresse93,kresse96,kresse96b,kresse99} The energy cut-off was set to 520 keV and the ${\bm k}$-spacing was set to $0.02\ $\AA$^{-1}$. The reciprocal lattice is sampled with a Monkhorst--Pack \cite{monkhorst76,monkhorst77} scheme for BCC Na and on a Gamma--centered scheme for all other materials. XC functionals and fast Fourier transformation (FFT) mesh grids used for plotting surfaces of a given charge density are listed in Table.~\ref{table2}.  Lattice parameters were obtained from experiments \cite{jain13} as the starting points for structural relaxation.

After optimizing the lattice parameters, we calculate the self-consistent charge density $\rho$ of the ground state from the Kohn--Sham orbitals:\cite{martin20}
\begin{equation}
\rho(\bm{r})=\frac{V_\Omega}{(2\pi)^3} \sum_{i,\bm{k}} f_{i,\bm{k}}|\phi_{i,\bm{k}}(\bm{r})|^2,
\end{equation}
\noindent
where $i$ runs over all particles, $\bm{k}$ runs over all $\bm{k}$ states, $\phi_{i,\bm{k}}$ the $i$th Kohn--Sham orbital at a $\bm{k}$ state, and the $f_{i,\bm{k}}$ are the corresponding occupation numbers. The Kohn--Sham orbitals are:\cite{martin20}
\begin{equation}
\phi_{i,\bm{k}}(\bm{r})=\frac{1}{\sqrt{n_\text{cell} }} \exp({\sf{i}}\bm{k}\cdot\bm{r})u_{i,\bm{k}}(\bm{r}),
\end{equation} 
\noindent
in which $n_{\text{cell}}$ is the number of unit cells in the crystal and the $u_{i,\bm{k}}$ are Bloch waves, given by:
\begin{equation}
u_{i,\bm{k}}(\bm{r})=\frac{1}{\sqrt{V_\Omega}} \sum_{m} c_{i,m}(\bm{k})\exp({\sf{i}}\bm{k}\cdot\bm{K}_m),
\end{equation}
\noindent
where the summation is over the reciprocal lattice vectors $\bm{K}_m$ required to calculate the charge density, whose number is indicated by the FFT mesh density.\cite{martin20} The coefficients $c_{i,m}(\bm{k})$ can be obtained by using a matrix equation based on the system Hamiltonian. Details are given in Ref.~\onlinecite{martin20}.  Our calculation is conducted over one unit cell, so we set $n_{\text{cell}}$ to one and the FFT grid density is the number of real-space grid points in the unit cell. As $\rho$ is a number density, we divide by the crystal volume here to obtain the dimension $[L^{-3}]$.

\subsection{Comparison between Charge Density and Minimal Surfaces}
\label{sec3B}

We use the root-mean-square (RMS) to estimate the deviations of the surfaces of constant charge density $S_\rho$ from its corresponding TPMS, denoted by $S$.  To calculate the RMS, we take a grid of $100\times100$ mesh points on each elementary piece of the unit cell of the TPMS, from which the entire unit cell can be generated from the operations of the point group. Thus, there are $8\times10^4$ points on the P surface, $1.6\times10^5$ points on the I-WP surface, $3.2\times10^5$ points on the F-RD surface, and $2.4\times10^5$ points on the H$^\prime$-T surface. $S_\rho$ is formed by $n_{ab}$ points on the FFT mesh with their corresponding charge density $\rho$ falling in an interval $[\rho_a,\rho_b]$. Thus, $\rho\approx{1\over2}(\rho_a+\rho_b)$. We then scale the TPMS, calculate the smallest distance $d_i$, $i\in n_{ab}$, between each point on $S_\rho$ and $S$, and establish a data set of the smallest distances, which we use to obtain the RMS.  The RMS is calculated as: 
\begin{equation}
{\rm RMS}=\sqrt{\sum_i^{n_{ab}}d_i^2\over n_{ab}}.
\end{equation}
Thus, the lower the dispersion of computed points from its TPMS, the lower the RMS, so the more $S_\rho$ resembles $S$.

\section{Results}
\label{sec4}

\subsection{Na}
\begin{figure}[t!]
\centering
\includegraphics[width=1\linewidth]{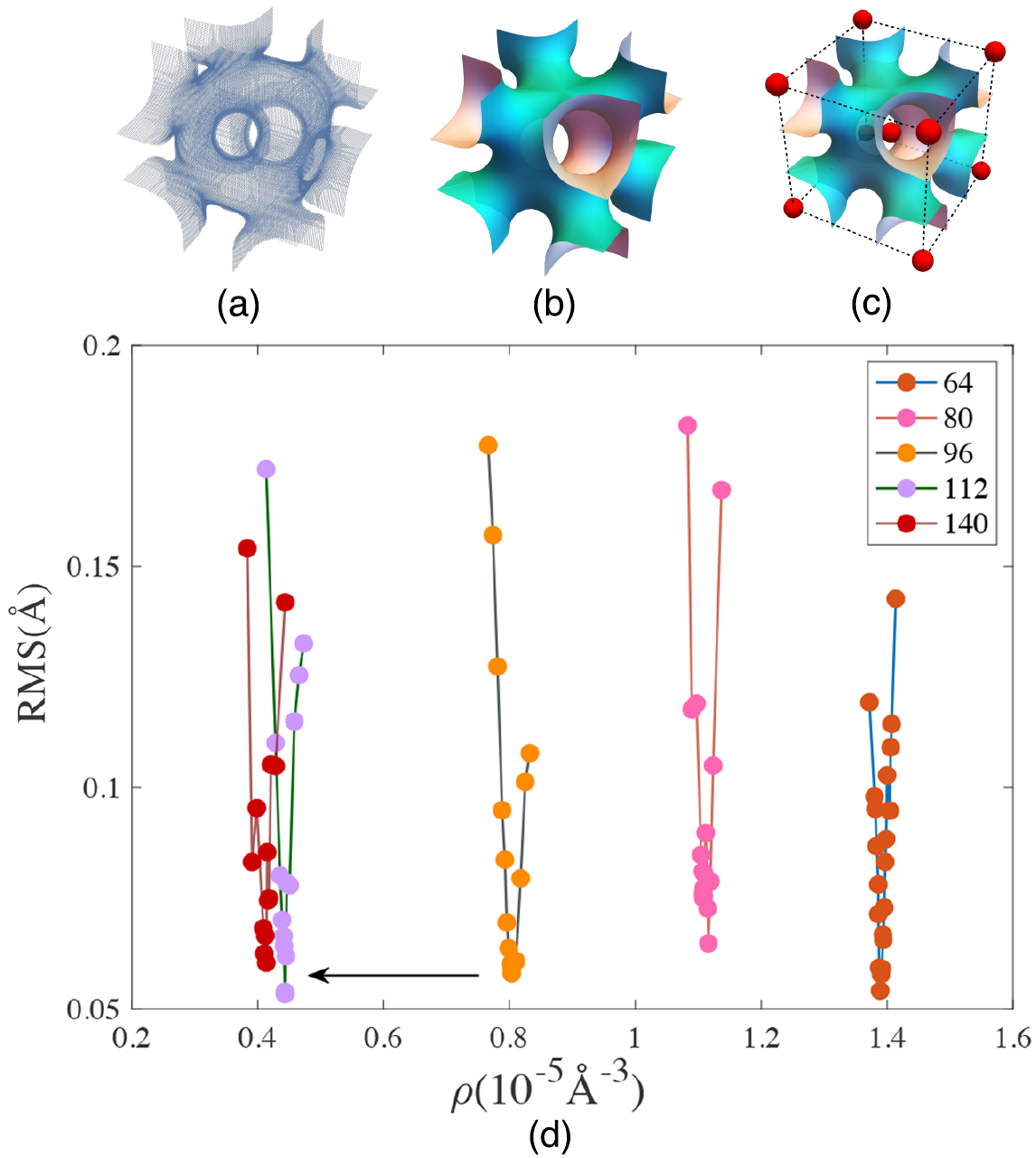}
\caption{(a) Charge density distributions of constant $\rho$ of BCC Na. Plot of constant $\rho$ at given values against (b) the I-WP surface. (c) Plot of the BCC lattice against the I-WP TPMS. Red spheres in (c) represent atoms. (d) RMS against charge density $\rho$ of Na in the BCC lattice. The points are where calculations were carried out and the lines are guides for the eye. The arrow points to the minimum of the RMS and the mesh densities are shown in the legend.}
\label{fig3}
\end{figure}

\begin{figure}[t!]
\centering
\includegraphics[width=1\linewidth]{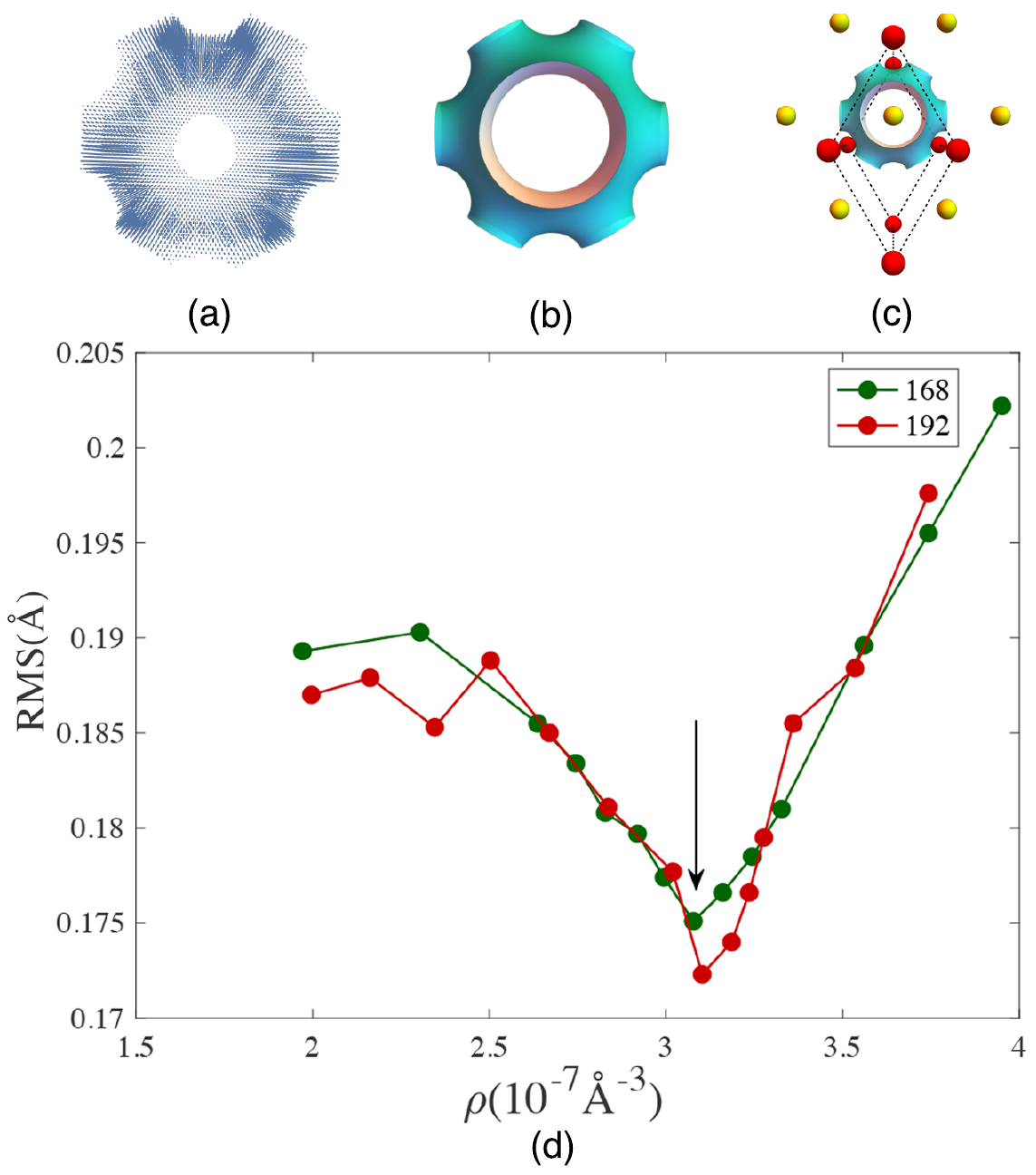}
\caption{(a) Charge density distributions of constant $\rho$ of HCP Na. Plot of constant $\rho$ at given values against (b) the H$'$-T surface. (c) Plot of the HCP lattice against the H$'$-T TPMS. Red and yellow spheres in (c) represent atoms on top/bottom and middle layers, respectively. (d) RMS against charge density $\rho$ of Na in the HCP lattice. The points are where calculations were carried out and the lines are guides for the eye. The arrow points to the minimum of the RMS and the mesh densities are shown in the legend.}
\label{fig4}
\end{figure}

Sodium (Na), an alkali metal, is highly reactive and rarely found in its pure state. Sodium has atomic number 11, with the electronic configuration $1s^2\,2s^2\,2p^6\,3s^1$. The half-filled $3s$-band is characterized by a spherical symmetry and can hybridize with the $3p$-band under some perturbations. The surfaces of constant energy of the $3s$-band remains nearly spherical until they are deformed close to the Brillouin zone boundary, showing its free-electron-like nature away from the Brillouin zone.\cite{ham62} Sodium exhibits a BCC structure (space group $Im\overline{3}m$) at room temperature under atmospheric pressure.  

Figure~\ref{fig3}(d) shows the RMS against the charge density for Na in the BCC lattice for the specified FFT mesh densities. The minimum decreases significantly with increasing mesh density and, for all meshes, the RMS shows sharp minima. Convergence with increasing mesh density is clearly observable. The converged minimum around $4.3\times10^{-6}\text{\AA}^{-3}$ is the best fit obtained. Figure~\ref{fig3}(a) shows the converged surface. The I-WP TPMS corresponding to the BCC lattice of Na is shown in Fig~\ref{fig3}(b) and the positions of the Na atoms against the I-WP surface is given in Fig.~\ref{fig3}(c). Thus, with the LDA, surfaces of zero charge density in BCC Na converge to the I-WP minimal surface. 

Sodium undergoes a martensitic transformation from BCC to HCP (space group $P6_3mm/c$), with a $c/a$ ratio close to the ideal of $1.63$, when cooled under 36K.\cite{barrett56,young75}  Calculations of the Helmholtz free energy based on DFT \cite{straub71} and the embedded-atom method (EAM) \cite{riffe24} confirm the martensitic transformation, though the EAM calculation reveal a more complex scenario of structural transitions. The convergence of HCP Na towards the H$^\prime$-T surface is shown in Fig.~\ref{fig4}(d), with the converged minimum near $3.1\times 10^{-7} \text{\AA}^{-3}$. The convergence to the minimum charge density is qualitatively different from that for BCC sodium, with the minimum occurring near the same point for different mesh densities. The converged surface is shown in Fig.~\ref{fig4}(a), the H$^\prime$-T surface shown in Fig.~\ref{fig4}(b), and the positions of the atoms on the TPMS indicated in Fig.~\ref{fig4}(c).

\begin{figure}[t!]
\centering
\includegraphics[width=1\linewidth]{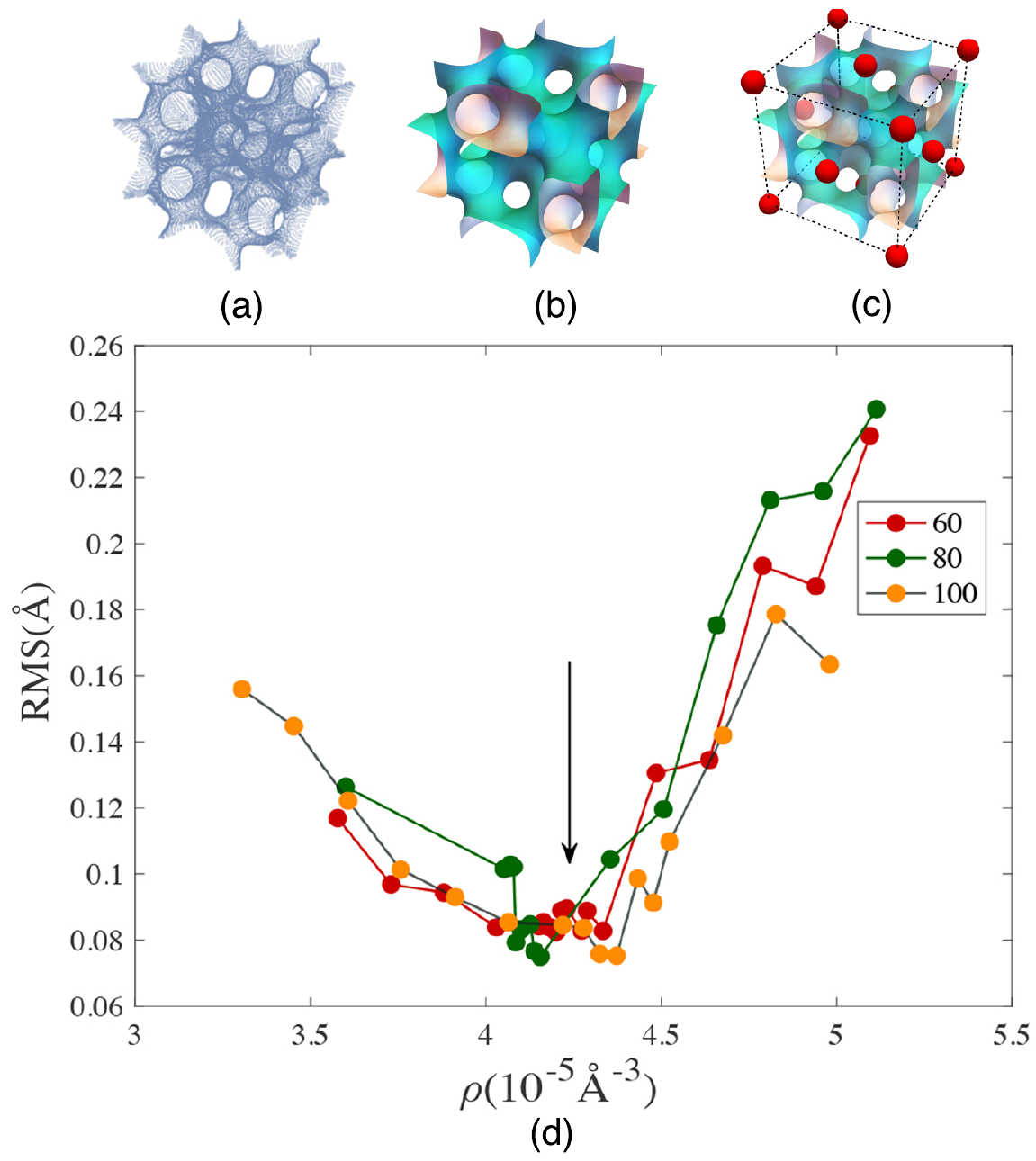}
\caption{(a) Charge density distributions of constant $\rho$ of FCC Al. Plot of constant $\rho$ at given values against (b) the F-RD surface. (c) plots the FCC lattice against the F-RD surface. Red spheres in (c) represent atoms. (d) RMS against charge density $\rho$ of Al in the FCC lattice. The points are where calculations were carried out and the lines are guides for the eye. The arrow points to the minimum of the RMS and the mesh densities are shown in the legend.}
\label{fig5}
\end{figure}

\begin{figure}[t!]
\centering
\includegraphics[width=1\linewidth]{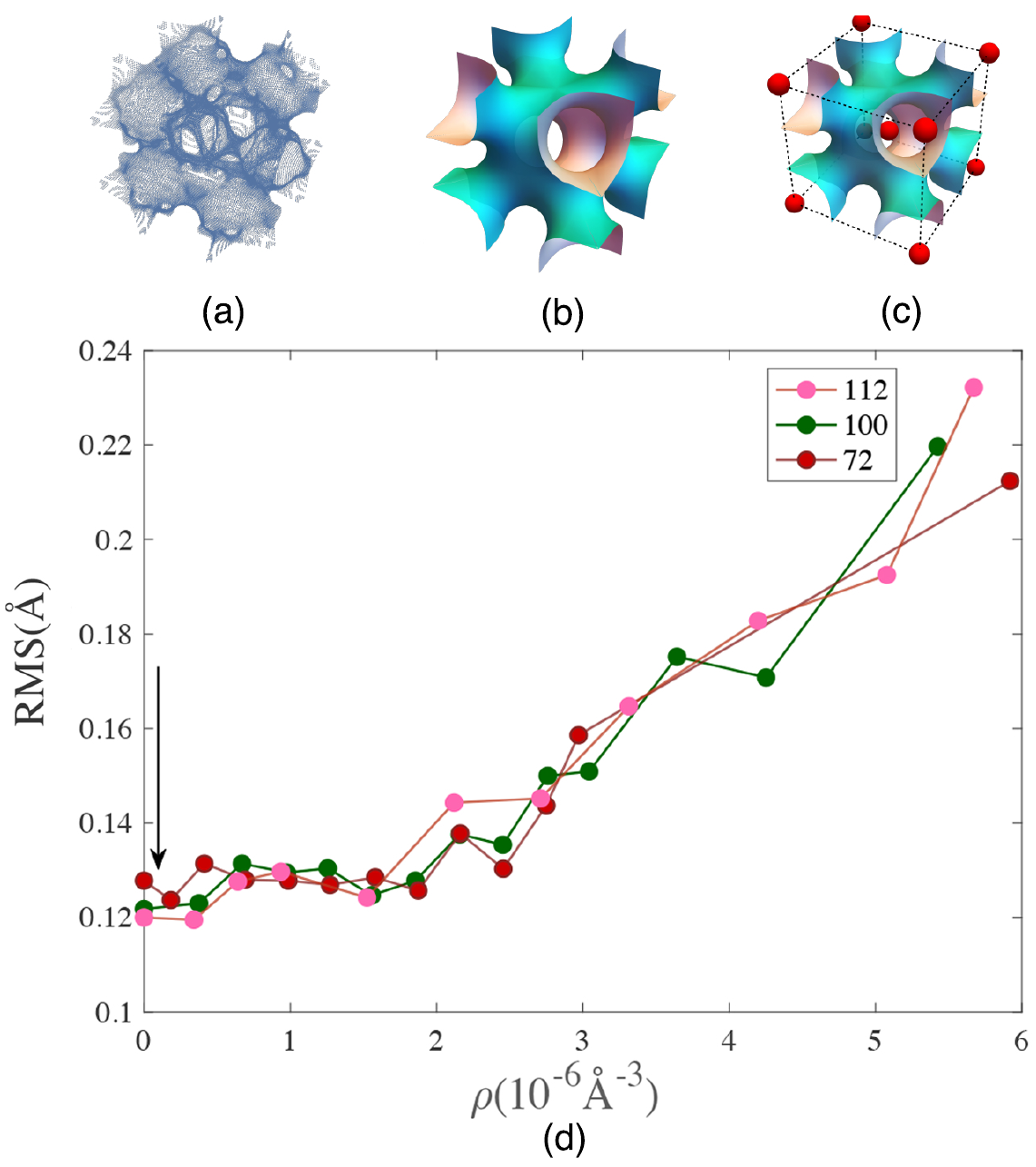}
\caption{(a) Charge density distributions of constant $\rho$ of BCC Al. Plot of constant $\rho$ at given values against (b) the I-WP surface. (c) plots the BCC lattice against the I-WP surface. Red spheres in (c) represent atoms. (d) RMS against charge density $\rho$ of Al in the BCC lattice. The points are where calculations were carried out and the lines are guides for the eye. The arrow points to the minimum of the RMS and the mesh densities are shown in the legend.}
\label{fig6}
\end{figure}

\subsection{Al}

\begin{figure}[t!]
\centering
\includegraphics[width=1\linewidth]{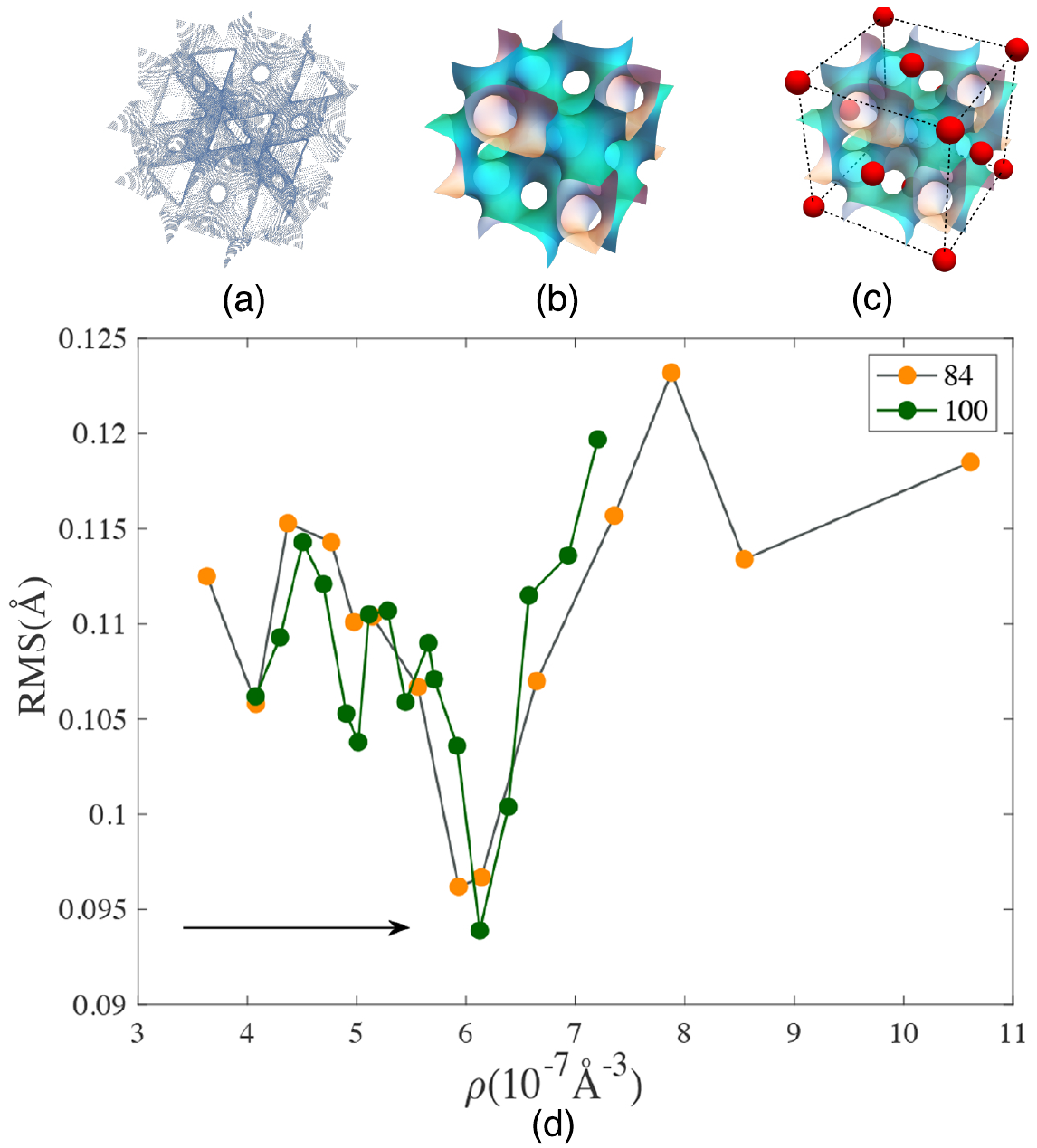}
\caption{(a) Charge density distributions for constant $\rho$ of FCC Cu. (b) The F-RD surface. (c) plots the FCC lattice against the F-RD TPMS. Red spheres in (c) represent atoms. (d) RMS against charge density $\rho$ of Cu in an FCC lattice. The points are where calculations were carried out and the lines are guides for the eye. The arrow points to the minimum of the RMS and the mesh densities are shown in the legend.}
\label{fig7}
\end{figure}

\begin{figure}[t!]
\centering
\includegraphics[width=1\linewidth]{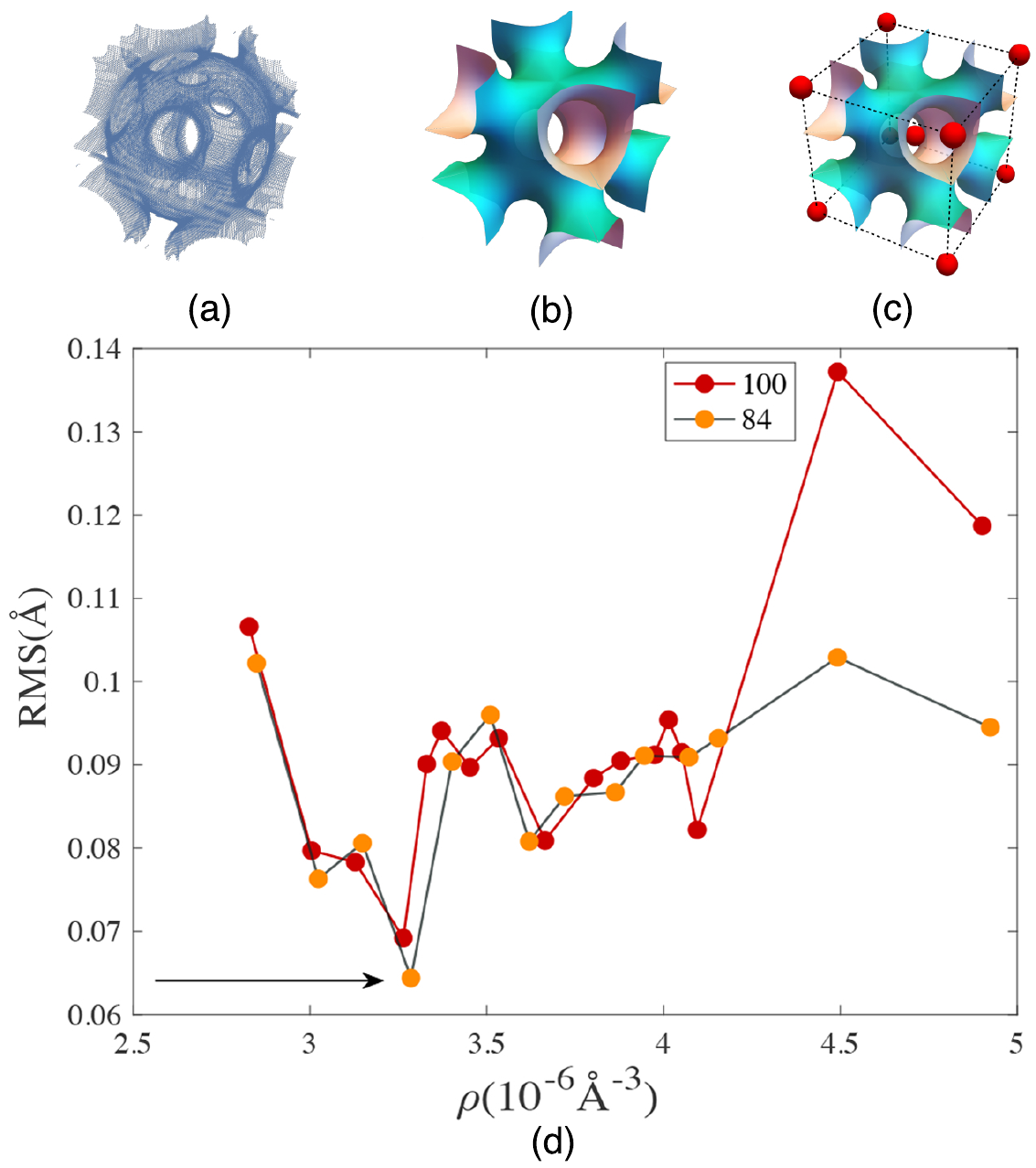}
\caption{(a) Charge density distributions for constant $\rho$ of BCC Cu. (b) The I-WP surface. (c) plots the BCC lattice against the I-WP TPMS. Red spheres in (c) represent atoms. (d) RMS against charge density $\rho$ of Cu in an BCC lattice. The points are where calculations were carried out and the lines are guides for the eye. The arrow points to the minimum of the RMS and the mesh densities are shown in the legend.}
\label{fig8}
\end{figure}

Aluminum (Al) is in the FCC lattice (space group $Fm\overline{3}m$) at standard temperature and pressure. With atomic number 13, aluminum has the electronic configuration [Ne]$3s^2\,3p^1$. The $3s$ and $3p$ orbitals hybridize to form $\sigma$ bonds that contribute to the lower parabolic-shaped bands below the Fermi energy, which strongly resemble bands of free electrons.\cite{segall61} The second and third $p$-bands are occupied primarily by the $3p$ valence electron, with some $3s$ contribution, and have $\pi$ character along the nearest-neighbor directions. The symmetry labels of the bands\cite{segall61} confirm this behavior.

Figure~\ref{fig5}(d) shows the RMS against the charge density for Al in the FCC lattice as a function of the indicated mesh density. The arrow points to the minimum value of the RMS. For all FFT meshes, the RMS shows sharp minima around $4.2\times10^{-5}$\AA$^{-3}$, with convergence with increasing mesh density and the minimum point shifted slightly to lower densities. The converged surface is shown in Fig.~\ref{fig5}(a). Figure~\ref{fig5}(b) shows the F-RD TPMS corresponding to the FCC Al and Fig.~\ref{fig5}(c) shows the positions of the Al atoms with respect to the TPMS. Thus, for FCC Al with the GGA-PBE XC potential, we see convergence of surfaces of a constant charge density to the F-RD surface as $\rho\to0$. 

Al has been shown to exhibit a BCC lattice by both a ramp-compression experiment at 475~GPa \cite{polsin17} and a synchrotron-based X-ray diffraction experiment. \cite{fiquet19} DFT calculations \cite{sjostrom16} show that FCC Al transforms to its BCC phase under 363~GPa at 920K. 

Convergence of surfaces of a constant charge density of Al in the BCC lattice towards the I-WP surface is given in Fig.~\ref{fig6}. The converged minimum is around $1\times10^{-7}\text{\AA}^{-3}$ (Fig.~\ref{fig6}(d)), with the converged charge density at or near $\rho=0$.  The RMS of the fit to the corresponding TPMS is near 0.12, which is somewhat higher than the ground state.  The best fit to the I-W surface, shown in Fig.~\ref{fig6}(a), s notable less resolved than that in Fig.~\ref{fig3}(b) for BCC Na.  This may be due to BCC Al not being a ground state crystal structure at high pressure, though other systems showing particular crystal structures at elevated temperatures do not show such deviations.

\subsection{Cu}

Copper (Cu) has the FCC lattice (space group $Fm\overline{3}m$) at standard temperature and pressure. It has atomic number 29 and an electronic configuration of [Ar]$3d^{10}\,4s^1$. The $4s$ bands are more dispersive compared with the $3d$ bands and can form wide bands in all high-symmetry directions. The $4s$ and $3d$ bands with the same symmetry will split near their initial crossing point due to the hybridization and the remaining unhybridized bands will form narrow bands across the Brillouin zone. \cite{segall62}

Figure~\ref{fig7}(d) plots the RMS against the charge density for FCC Cu as a function of the indicated mesh densities. For all meshes, the RMS shows sharp minima around $6.1\times 10^{-7}\text{\AA}^{-3}$, with convergence increasing slightly with the increasing mesh density. The converged surface, the F-RD surface and the positions of the Cu atoms with on the TPMS, are shown in Fig.~\ref{fig7}(a,b,c). Thus, for FCC Cu with the GGA-PBE XC potential, there is a convergence of surfaces of constant charge density to the F-RD surface.

The BCC phase of copper has been proposed and observed in several experimental scenarios. The bulk phase has been seen in shock-compression experiments, \cite{sims22} where it was concluded that this phase is stabilized at high temperatures and remains stable at high pressures.  The BCC phase of copper was also proposed and observed by high-resolution transmission electron microscopy to grow into a thin layer at a tilt grain boundary. \cite{schmidt95,schmidt98} DFT calculations \cite{alfke22} predict that the BCC phase is more stable for small particles, where surface energy is dominant, while the bulk energy becomes more important upon particle growth, which favors the FCC structure. Figure~\ref{fig8}(d) shows the RMS against the charge density for BCC Cu for two mesh densities. The convergence towards the I-WP surface in Fig.~\ref{fig8}(a) shows a deep minimum near $3.3\times10^{-6}$\AA$^{-3}$.

\subsection{Zr}

Zirconium (Zr) is a transition-metal with atomic number 40 that crystallizes in an HCP lattice (space group $P6_3mm/c$) at standard pressure and temperature. The valence electrons of Zr are $4d^2\,5s^2$.  The electronic structure of Zr along the principal symmetry directions \cite{iyakutti76} consists of a broad $5s$ band, which hybridizes with the $4d$ band of the same symmetry, and the Fermi level passes through $4d$ and $5s$ bands. There are two martensite phases of zirconium:~the BCC phase (space group $Im\overline{3}m$) at high temperature, and the hexagonal phase (space group $P6/mmm$) under high pressure. \cite{hao08} 
 
 \begin{figure}[t!]
 	\centering
 	\includegraphics[width=1\linewidth]{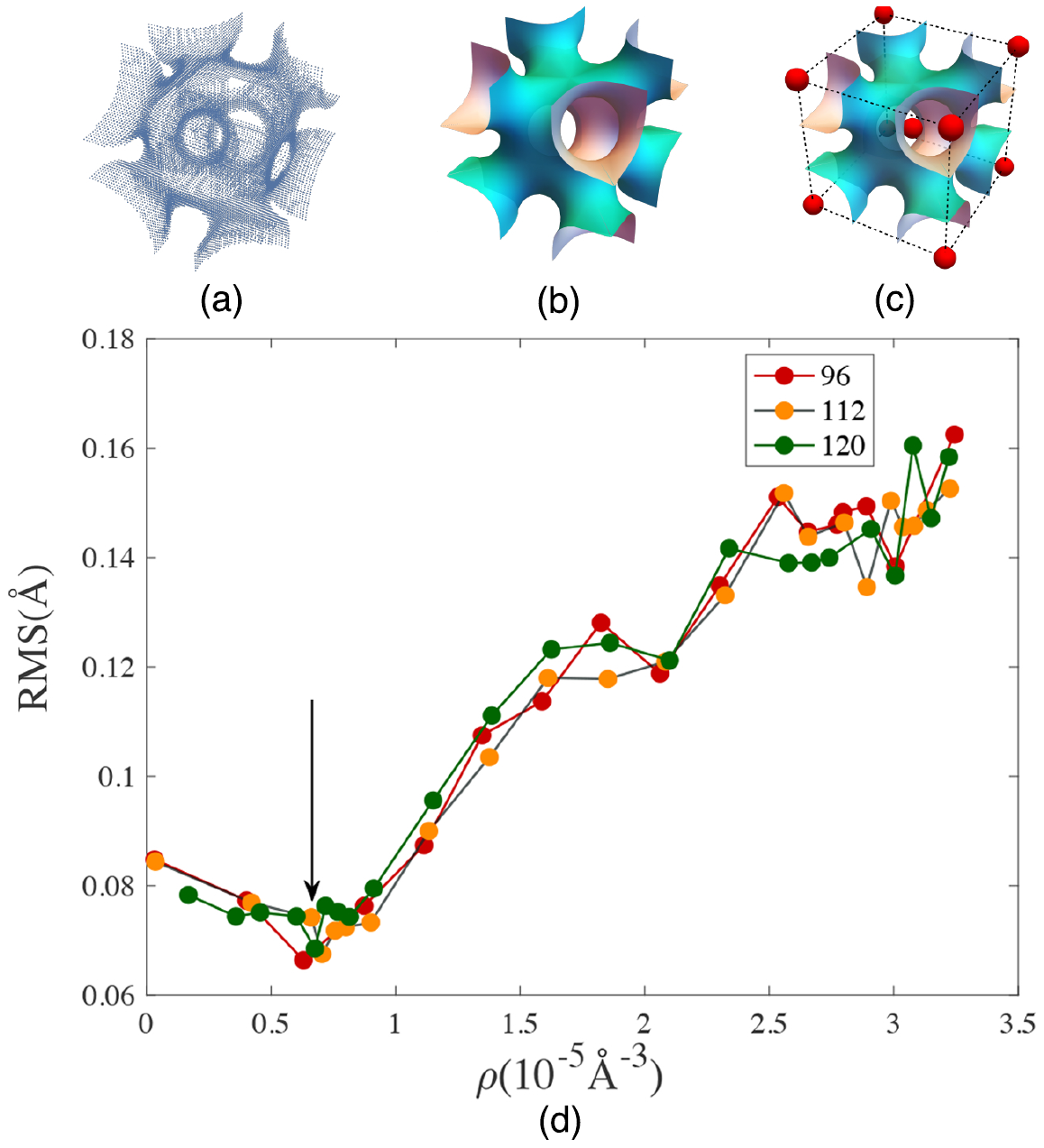}
 	\caption{(a) Charge density distributions of constant $\rho$ of BCC Zr ($Im\overline{3}m$). (b) The I-WP surface. (c) sketches atoms in the BCC unit cells against the I-WP surface. Red and yellow spheres represent Zr atoms. (d) RMS against charge density $\rho$ of Zr in a BCC lattice.  The points are where calculations were carried out and the lines are guides for the eye. The arrow points to the minimum of the RMS and the mesh densities are shown in the legend.}
 	\label{fig9}
 \end{figure}

 \begin{figure}[t!]
 	\centering
 	\includegraphics[width=1\linewidth]{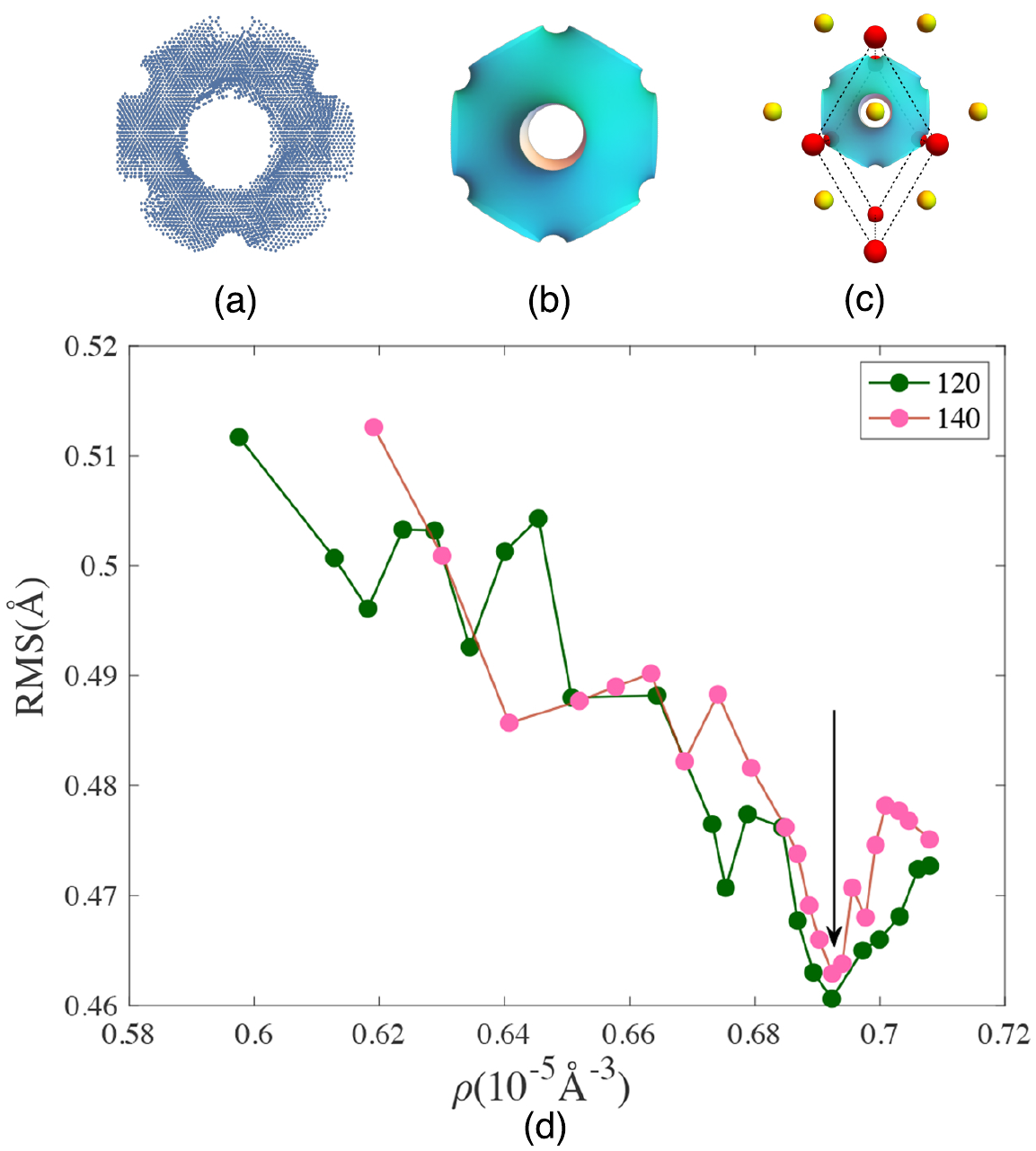}
 	\caption{(a) Charge density distributions of constant $\rho$ of HCP Zr ($P6_3/mmc$). (b) The H$^\prime$-T surface for the HCP Zr. (c) sketches atoms in the HCP unit cells against the H$^\prime$-T surface. Red and yellow spheres represent Zr atoms on top/bottom and middle layers, respectively. (d) RMS against charge density $\rho$ of Zr in the HCP lattice. The points are where calculations were carried out and the lines are guides for the eye. The arrow points to the minimum of the RMS and the mesh densities are shown in the legend.}
 	\label{fig10}
 \end{figure}

 \begin{figure}[t!]
 	\centering
 	\includegraphics[width=1\linewidth]{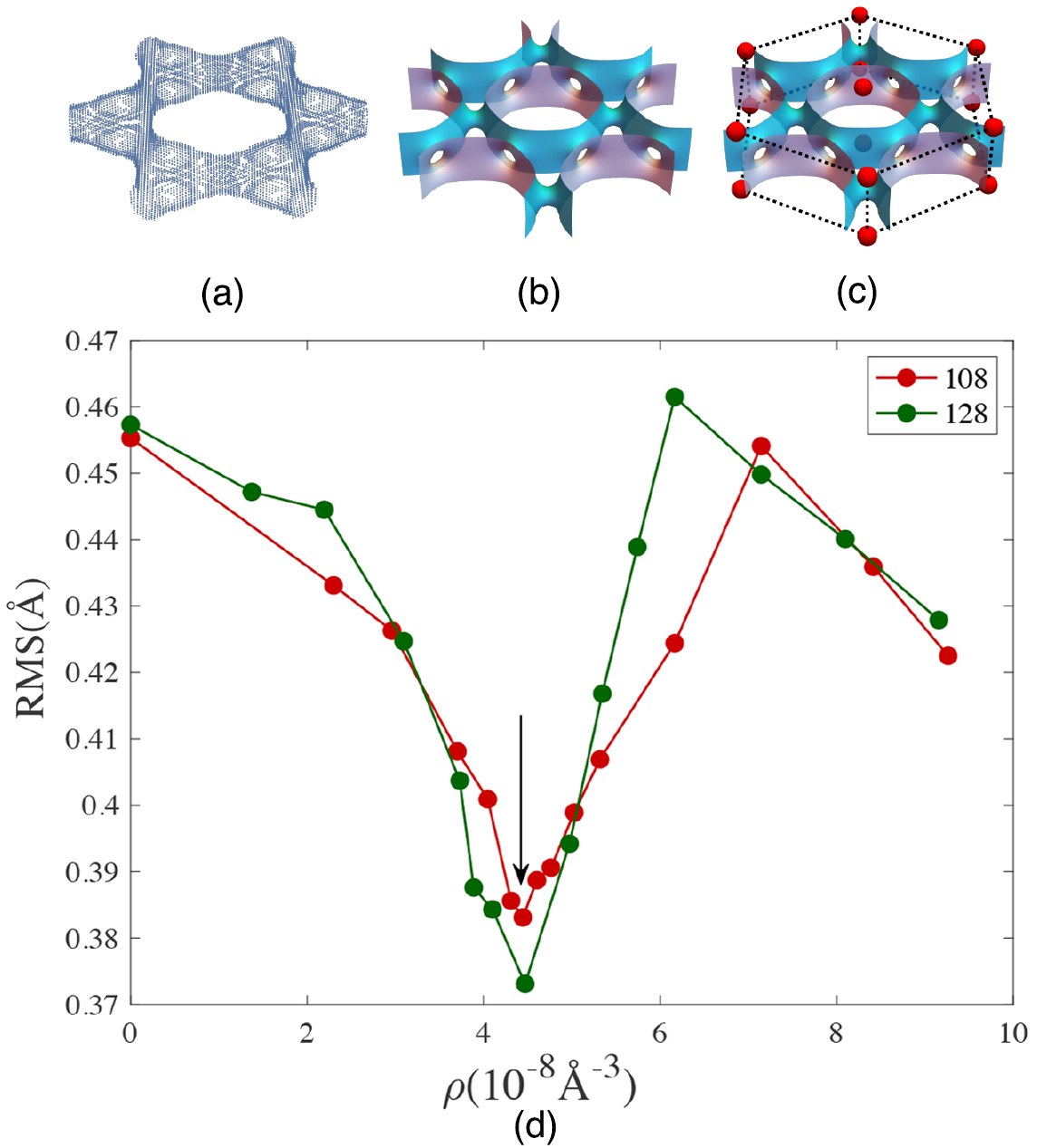}
 	\caption{(a) Charge density distributions of constant $\rho$ of hexagonal Zr ($P6/mmm$). (b) The H$^\prime$-T surface for the hexagonal Zr. (c) sketches atoms in the hexagonal unit cells against the H$^\prime$-T surface. Red and yellow spheres represent Zr atoms on different layers. (d) RMS against charge density $\rho$ of Zr in the hexagonal lattice. The points are where calculations were carried out and the lines are guides for the eye. The arrow points to the minimum of the RMS and the mesh densities are shown in the legend.}
 	\label{fig11}
 \end{figure}

Figure~\ref{fig9}(d) shows the RMS against the charge density for Zr in the BCC lattice as a function of the indicated mesh density. For all FFT meshes, the RMS decreases towards their minima, with convergence to the I-WP TPMS appears near $6.8\times10^{-6}$\AA$^{-3}$. Figure~\ref{fig9}(a) shows the converged surface, the I-WP TPMS is shown in Fig.~\ref{fig9}(b), and atom positions of the BCC zirconium with respect to the I-WP surface are shown in Fig.~\ref{fig9}(c). Therefore, for BCC Zr with the LDA XC potential, a convergence of the surfaces of constant charge density to the I-WP minimal surface is observed. 

Figure~\ref{fig10}(d) plots the RMS against the charge density for Zr in the HCP lattice as a function of the indicated mesh density. The RMS decreases towards to their minima with the convergence to the H$^\prime$-T surface near $6.9\times 10^{-6}\text{\AA}^{-3}$. The converged surface, the H$^\prime$-T surface and atom positions of the HCP zirconium with respect to the H$^\prime$-T surface are given in Fig.~\ref{fig10}(a), Fig.~\ref{fig10}(b) and Fig.~\ref{fig10}(c), respectively. Hence, for HCP Zr with the PBE XC potential, a convergence of the surfaces of constant charge density to the H$^\prime$-T minimal surface is observed. Convergence of Zr in the hexagonal lattice towards the H$^\prime$-T surface is given in Fig.~\ref{fig11}, with the convergence minimum locating around $4.47\times 10^{-8}$\AA$^{-3}$.

\subsection{NiTi}

The discovery of NiTi \cite{buehler63} has attracted attention for being a material that recovers from an initial plastic deformation in a low-temperature martensite phase to its original shape upon heating above a transformation temperature to a austenite phase. This so-called shape memory effect has been known since the 1930s in AuCd alloys. \cite{olander32} However, studies of this effect began in earnest with the discovery of Ni-Ti alloys around 1960 because the shape memory effect is combined with superelasticity, corrosion resistance, biocompatibility, and superior engineering properties that has seen many applications, for example,  as stents and implants. \cite{patel20} NiTi has also stimulated DFT studies on the pathways between austenite and martensite. \cite{hatcher09,hatcher09b}

\begin{figure}[t!]
\centering
\includegraphics[width=1.\linewidth]{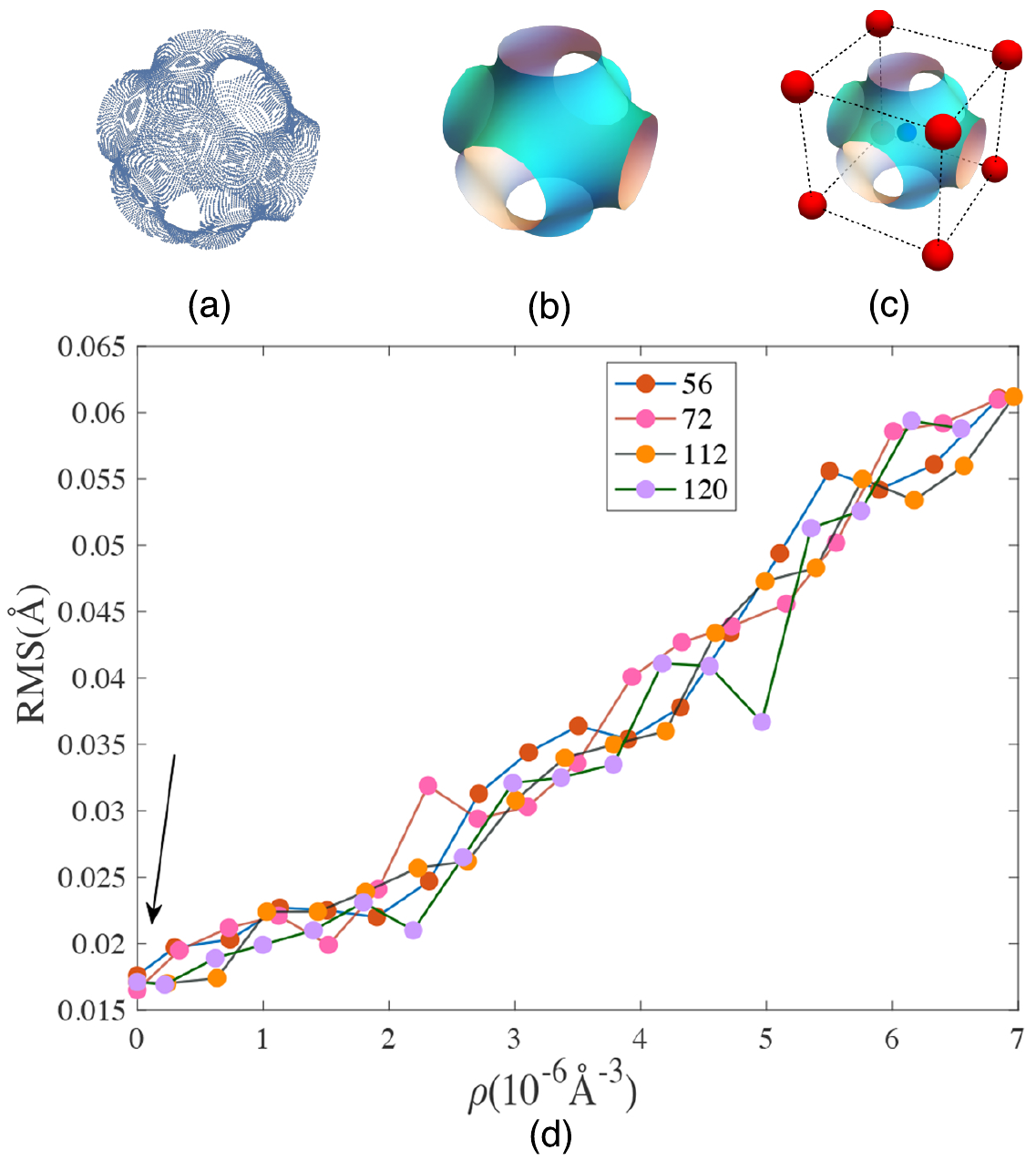}
\caption{ (a) Surfaces of constant charge density of NiTi ($Pm\overline{3}m$). (b) The $P$ surface for B2 NiTi. (c) plots the SC lattice against the P surface. Red balls and blue ball in (c) represent real atoms. (d) RMS shown against charge density $\rho$ of NiTi in the SC lattice. The points are where calculations were carried out and the lines are guides for the eye. The arrow points to the minimum of the RMS and the mesh densities are shown in the legend.}
\label{fig12}
\end{figure}

The structure of NiTi in the austenite (B2) phase has the $Pm\overline{3}m$ space group, which is simple cubic (SC) with two atoms in the unit cell [Fig.~\ref{fig12}(c)], the same as the CsCl structure, with one type of atom at the corners of the cubic unit cell and the other type at the center. For this reason, the structure of the austenite phase of NiTi is often (mistakenly) called a BCC structure. 

The RMS against the charge density for B2 NiTi as a function of the indicated mesh densities is shown in Fig.~\ref{fig12}(d). For all mesh densities the RMS converges uniformly toward zero, though there is no clearly observed minimum because the RMS decreases to its minimum value as the charge density decreases.  The best fit to the minimal surface obtained by the charge density is shown in Fig.~\ref{fig12}(a) and the P surface in Fig.~\ref{fig12}(b). The atomic positions in NiTi are shown in Fig.~\ref{fig12}(c). Note that the atomic positions do not correspond to lattice points, which are located at flat points of the minimal surface. \cite{yin24}

\section{Discussion}
\label{sec6}

\begin{figure}[b!]
	\centering
	\includegraphics[width=.8\linewidth]{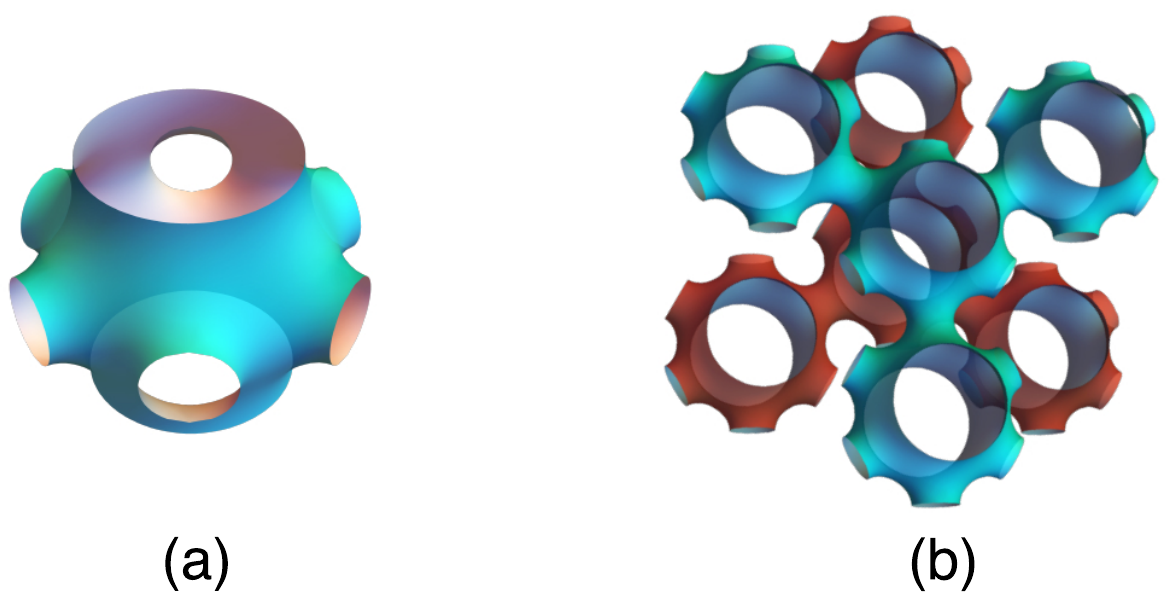}
	\caption{(a) The smallest repetition unit of the H$'$-T surface. (b) The repetition unit chosen that can realize the sub-lattice translation required by the $P6_3/mmc$ space group symmetry. In (b), surfaces in different colors are in different layers.}
	\label{fig13}
\end{figure}

We have determined the zero-charge density surfaces obtained by introducing exchange and correlation into electronic interactions by using DFT. This generalizes the approach whereby TPMS are obtained from the electrostatic field determined by an arrangement of static electric charges that has been the subject of previous work.\cite{schnering87,schnering91,mackay93,gandy01,gandy02} We have used the RMS to find the best fit to TPMS associated with various materials. We studied Na, Al, Cu, Zr and NiTi and show that their surfaces of given charge densities converge to TPMS when the charge density goes towards zero, such that the point group of the crystal is a subgroup (proper or improper) of the point group of its corresponding TPMS. Moreover, such a point group relation shows that:~(i)) a compatibility between the TPMS and the Bravais lattice of the crystal, and (ii) the periodicity of the TPMS conforms to the periodicity of the  crystal given by its corresponding lattice vectors. For the space group, the TPMS either is in the same space group, or can be modified to exhibit the same space group as its corresponding crystal. To be specific, the HCP Zr in Table.~\ref{tab1} is in the space group $P6_3/mmc$ while the H$'$-T surface is in the space group $P6/mmm$. However, $P6_3/mmc$ symmetry can be realized in the H$'$-T surface by choosing the repetition unit to be a combination of 8 copies of the smallest repetition unit having $D_{6h}$ point group symmetry, as shown in Fig.~\ref{fig13}. 

Our calculations provide fundamental physical realizations of these mathematically proposed surfaces, including electronic characterizations of crystalline materials, rather than simply as structural models. Moreover, if we compare the genus of TPMS listed in Table.~\ref{table2}, we see that materials undergoing a martensitic phase transformation have their end phases corresponding to TPMS with the same genus. More detailed discussions on this point will be presented elsewhere. 

Minimal surfaces have been used to describe many classes of materials, \cite{hyde96} but the first to recognize the importance of TPMS for solid state transformations were Hyde and Andersson. \cite{hyde85,hyde86} These authors associated TPMS with the initial and final states of a martensitic transformation. Although they were interested in iron, the same ideas can be applied to other martensitic materials and to the shape memory effect. The positions of lattice points during a martensitic phase transformation, including twinning, is determined by the TPMS calculated from the charge density, which will affect the barriers to the transformation. We have discussed some of these ideas in Ref.~\onlinecite{yin24}. The main point is that we have associated TPMS with an observable quantity (the charge density) which will enable the interpretation of the topological interpretation of martensitic transformations in physical terms.

\end{document}